# A computational framework for the near-elimination of spreadsheet risk

Dr. Yusuf Jafry (Chief Technology Officer), Fredrika Sidoroff, Roger Chi
Risk Integrated
Third floor, Central Chambers
6 Victoria Street
Douglas, Isle of Man
IM1 2LH, British Isles
Yusuf.Jafry@RiskIntegrated.com

**ABSTRACT**

*We present Risk Integrated's Enterprise Spreadsheet Platform (ESP), a technical approach to the near-elimination of spreadsheet risk in the enterprise computing environment, whilst maintaining the full flexibility of spreadsheets for modelling complex financial structures and processes. In its Basic Mode of use, the system comprises a secure and robust centralised spreadsheet management framework. In Advanced Mode, the system can be viewed as a robust computational framework whereby users can "submit jobs" to the spreadsheet, and retrieve the results from the computations, but with no direct access to the underlying spreadsheet. An example application, Monte Carlo simulation, is presented to highlight the benefits of this approach with regard to mitigating spreadsheet risk in complex, mission-critical, financial calculations.*

**1. INTRODUCTION**

Spreadsheet risk is the danger that errors in a common business tool such as Microsoft Excel can cause material losses when used inappropriately by financial organisations. Today most banks already have established sets of rules, standards, and controls over their accounting systems and many of the databases they access. However controls have not yet been put into place for their smaller systems such as spreadsheets. This lack of internal control and audit reporting is something now being addressed in response to the Sarbanes-Oxley Act of 2002 banking regulation in the US [SOX], and the operational risk section of Basel II globally [Basel II].

The problem with spreadsheets is that they are being built, in general, by non-programmers. Errors creep into formulas, some from formatting, and some from links to other spreadsheets. Most are due to negligence, although a few are due to fraud. There are no thorough procedures in place to check the accuracy of the spreadsheets, or to test multiple runs of data through them.

The financial software industry has responded in a number of ways. For example, there are various spreadsheet auditing tools available that can assist in the process of building reliable spreadsheets; and there are tools available to help end-users build new spreadsheets from scratch in the most robust manner. Also, more comprehensive solutions for managing the use of spreadsheets are beginning to emerge in the marketplace. These so-called Business Intelligence (BI) platforms are aimed at delivering insight to managers from masses of quantitative data centrally-held in the enterprise. For example, Microsoft is reportedly working on a BI solution around a future release of their Excel spreadsheet product, due for release in 2007 [The Banker, 2006].

In Section 2, we present Risk Integrated's technical approach to the near-elimination of spreadsheet risk in the enterprise computing environment whilst maintaining the full flexibility of spreadsheets for modelling complex financial structures and processes. The approach is particularly suited to those applications where a common set of complex spreadsheet-based calculations has to be applied across multiple instances of data inputs e.g. individual deals within a portfolio, distributed across the enterprise, whilst retaining centrally-managed consistency and integrity throughout.

By way of illustrative example, Section 3 discusses the application of the approach to the computationally-intensive process of Monte Carlo simulation for assessment of credit risk (another topical subject in the sphere of Basel II).

Section 4 contains concluding remarks.

## 2. RISK INTEGRATED'S ENTERPRISE SPREADSHEET PLATFORM

### 2.1 "Basic Mode": a secure spreadsheet management system

As depicted in Figure 1, Risk Integrated's Enterprise Spreadsheet Platform (ESP) represents a secure spreadsheet management framework whereby the spreadsheets themselves are exposed only to a select few designated experts ("superusers") in the organisation. These users have the responsibility for maintaining the integrity of the spreadsheet models, and for "uploading" their tested and "signed-off" versions to a centralised server/database. ESP provides sophisticated tools that monitor, assign, and track changes to those spreadsheets. A full audit trail is available for tracking the versioning of the spreadsheet models against those users submitting the changes, thereby eliminating that source of spreadsheet risk associated with the proliferation of un-versioned models, scattered around the organization.

The employment of ESP in the manner described, i.e. as a secure and robust spreadsheet management system, is considered the Basic Mode of ESP operation, and may be considered as comparable with the aims of the other emerging Business Intelligence platforms with regards securing access to spreadsheets.

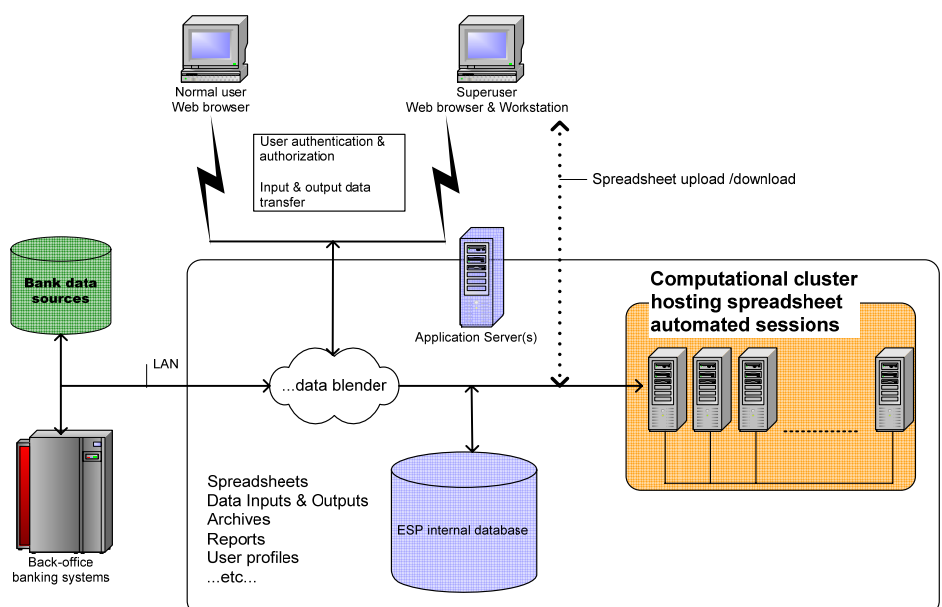

Figure 1. Risk Integrated's Enterprise Spreadsheet Platform (ESP) architecture.

**2.2 "Advanced Mode": a secure & robust computational engine**

The full power of ESP is realised in the Advanced Mode of ESP operation whereby normal end-users (in contrast to superusers) can utilize a given spreadsheet model as a computational engine, but without having direct access to the underlying spreadsheet itself. The rationale here is that those users whose primary goal is to input data and make use of results of computations do not need access to the underlying spreadsheets. In other words, they can only submit "jobs", via their web-browser window, to a centralised cluster of computational servers, which, in turn, retrieve the spreadsheets via the management framework, automatically spawn service sessions of the underlying spreadsheet application, and "run" the computations against the "inputs" submitted by the end-users. The spreadsheets never reside on the end-user (client) machines and are never opened interactively by the end-users, thereby eliminating the major source of spreadsheet risk caused by end-users, namely, the introduction (inadvertently or otherwise) and propagation of errors within the core of the spreadsheets themselves. Depending on the specificity of the spreadsheet in question, the data submitted by the end-users via their web-browser can be pre-screened (before being sent to the computational servers) using field-wise validation technology built-in to the web application interface, thereby minimising another source of spreadsheet risk, namely "nonsense" data being fed to the computations.

The results of the computations are time-stamped and archived in the central database (against the given set of end-user inputs and spreadsheet model version) for auditability and reporting, before being sent back to the end-users' browser for display. This minimises another source of spreadsheet risk, namely the manipulation of the results coming out of the spreadsheet calculations.

**2.3 Security layer**

The security of the ESP management framework leverages the authentication and authorisation layers inherent to the operating system and/or Relational Database Management System (RDBMS). ESP therefore allows for completely configurable user and group level security and permissions.

**2.4 User and Data Interfaces**

A possible criticism of the approach pertains to the perceived limited flexibility of the user interface for communicating with the underlying spreadsheet. One of the major reasons why spreadsheets have become so prevalent is because of their extreme ease-of-use, particularly with regard to rapid prototyping of ideas. To diminish this ease-of-use and flexibility would undermine the use of the spreadsheet format.

However, the ESP, by definition, *retains* that flexibility for the appropriate users in the organisation (namely the designated experts) by providing them with complete access to the underlying spreadsheets (albeit through a secure content-management layer). They can prototype, manipulate, and test the spreadsheets in the normal manner. By contrast, the normal end-users, who, in order to minimize operating risk, do not have access to the underlying spreadsheets, are provided with non-programmable Graphical-User-Interfaces (GUIs) which enable them to pass data in to the spreadsheet, perform the desired calculations, and retrieve the results back. For maximum flexibility, these interfaces can be generic. For example, they may comprise a simple suite of data-entry grids which map on to the corresponding "inputs worksheet(s)" of the underlying spreadsheet, plus a corresponding suite of output data fields mapped on to the respective "outputs worksheet(s)" of the underlying spreadsheet. Alternatively, for well-established spreadsheet models (i.e. those which are used as deployed "applications" rather than prototyping scratchpads), the GUI's can be customised to reflect the model-specific input and output fields (thereby facilitating the use of field-wise validation technology built-in to the web application interface mentioned earlier, thereby further minimising spreadsheet risk).

With ESP's implicit separation between the data and the spreadsheets, data can be fed to and from the spreadsheets in a variety of ways in addition to the generic (or customized) user GUIs discussed above. For example, it is straightforward to automatically populate the spreadsheets via links to the company's existing banking systems and data sources. This eliminates the spreadsheet risk associated with manual "double-entry" and/or "copy-and-paste" of data from the data sources into the spreadsheets.

The outputs are similarly flexible. They can be displayed to the user, exported to a variety of formats, or used to re-populate a banking system, database, or data warehouse (with the inherent advantages of consolidated reports, etc).

**2.5 Possible usage scenarios**

The ESP can accommodate any spreadsheet for any purpose. For example, in a typical Basic Mode usage scenario, it would simply be employed as a secure spreadsheet management system where users are all, by definition, "superusers" i.e. they use the spreadsheets by first downloading them from the central store, manipulate them for their present purpose, then upload them again for safe keeping (plus archiving and version control, etc). Even this basic usage scenario dramatically reduces spreadsheet risk by eliminating the proliferation of models scattered across users' desktops. However, it does not take advantage of the full capabilities of ESP in Advanced Mode. By contrast, the example discussed below in Section 3, maximally illustrates the use of the ESP as a secure and robust computational engine, in addition to its spreadsheet-management role.

**2.6 Domain and Mechanical errors**

Ultimately the weak link in the chain is with the superusers. Domain (i.e. incorrect specification) and/or mechanical (i.e. incorrect implementation) errors introduced by them (unintentionally or otherwise) into the underlying spreadsheets can adversely affect the computations. However, ESP assists in these areas too.

**Specialisation**

As discussed in the example presented in Section 3, ESP enables applications to be developed such that only the core business logic is programmed in the spreadsheet by the business analyst. All other aspects (e.g. data handling, numerical algorithms, etc) can be embodied within the ESP computational framework, which can be independently tested and qualified as fit-for-use. In this way, the business analysts only need to "program" (i.e. build spreadsheets) in their areas of expertise (i.e. business logic), thereby minimising the chance of them introducing errors outside of their area of core specialisation. In this way, ESP still allows business users to create their own applications (i.e., spreadsheets encompassing business logic) and, therefore, "avoid much of the IT bottleneck" [Gartner, 2006] that would be incurred with software development.

**Testability**

However, errors do occur, and they can only be reliably identified and removed through a rigorous testing process. ESP provides for this in a systematic way. Since the data layer is separated from the spreadsheet core, it is straightforward to establish a set of "standard" tests comprising a collection of input/output data sets which have been agreed and signed-off as being valid and correct. Thereafter, whenever a spreadsheet is modified by a superuser, the policy can be imposed that the spreadsheet can only "go live" when it has successfully passed the battery of standard tests. Moreover, as discussed in Section 3, for any calculations involving pseudo-random numbers, the issue of testability becomes more severe. ESP enables full control of the pseudo-random "seeds" such that input/output data sets can be fully replicated (in contrast, for example, to Excel, which does not allow such control).

**Spreadsheet Audit Trail**

In the event that the effects of domain and/or mechanical errors do creep in to the core spreadsheets, the auditability inherent to ESP can assist in resolving the issue. All modifications to the spreadsheet models are logged (with user ID's of those making the modifications) by the ESP content-management system. Moreover, any time a spreadsheet model is used by the system at runtime, a copy is automatically made and archived on the server against the given usage instance, thereby providing a secure electronic audit trail back to the actual model used, and the ID of the superuser who last modified the model. It would require collusion between the superuser, database administrator, system administrator, security officer, and network administrator to tamper with this audit-trail.

## 3. EXAMPLE APPLICATION: MONTE CARLO SIMULATION

### 3.1 Introduction

Monte Carlo simulation is a well-established technique for analysing credit risk. As illustrated in Figure 2, the basic idea is to construct a (typically time-domain) cashflow model which captures the logic of the given deal, drive this model "into the future" with a set of randomised macro-economic scenarios (albeit with historically-imposed correlations between the variables), then perform the risk analyses on the outputs generated from all of the scenarios. The modelling process can be complex (e.g. if analysing a complicated energy-generation deal or a large commercial real estate deal with many properties and related securities) and the computational process can be numerically intensive, especially if globally-linked economies have to be included (e.g. with hundreds of interacting macro-economic variables from many different geographies) over long lifetimes (e.g. twenty-year deal lifecycles).

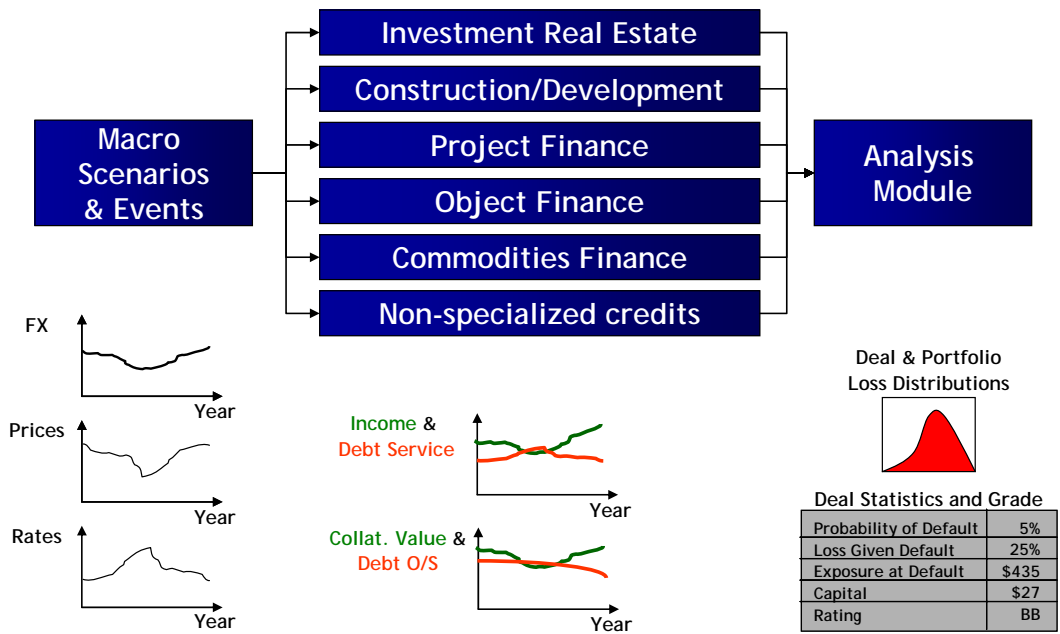

Figure 2. The basic principle of Monte Carlo simulation for analysing credit risk is illustrated. The simulation framework (macro-economic scenario generation and data analysis) is common to all asset-classes, and the cashflow model is tailored to a specific asset class (e.g. Investment Real Estate, Project Finance, etc., as shown) or indeed, to an individual deal.

**3.2 Spreadsheet or "black box"?**

A central issue faced by banks when constructing such Monte Carlo simulations is that the cashflow models are invariably constructed in a spreadsheet format (typically Microsoft Excel) either by their own business analysts (whose programming capabilities extend only to Excel), or by the banks' clients (or other external stakeholders in the deal) who present the Excel cashflow model as part of the deal "documentation". Under these (highly typical) circumstances, there are essentially two options available for performing the Monte Carlo simulation: i) the cashflow model is retained in Excel, and the rest of the simulation framework is built in Excel and/or Visual-Basic-for-Applications (VBA) "add-ins" etc; or ii) the cashflow model is re-written (by a software developer) in a different language (e.g. C++ etc) and then integrated into a non-spreadsheet-based simulation framework.

The first option—to build the entire simulation framework in Excel around the cashflow model—has the disadvantage of potentially incurring severe spreadsheet risk – not least because the programming challenge to construct a mathematically-consistent and numerically-robust Monte Carlo simulation framework is decidedly non-trivial, beyond the skills of typical business analysts. Moreover, once such a framework has been built in spreadsheet form, it is highly-susceptible to the familiar forms of spreadsheet risk since the spreadsheet will, by its very nature, be large and complex, and thus vulnerable to the introduction of errors (inadvertent or otherwise).

The latter option, i.e. to translate the Excel cashflow model into another programming language (for further integration into a non-spreadsheet simulation framework) has the severe disadvantage that the underlying model can no longer be manipulated by the business analyst. The model has become the proverbial "black box", and all changes have to be implemented by a software developer.

**3.3 The ESP solution**

Using ESP in Advanced Mode, the advantages of both options can be realised, and the disadvantages of each eliminated. Specifically, as illustrated in Figure 3, ESP incorporates a code layer which enables the spreadsheet to be embedded within a robust computational engine (written in C++) which communicates with an auto-spawned session of the spreadsheet application (Microsoft Excel) via a shared-memory interface.

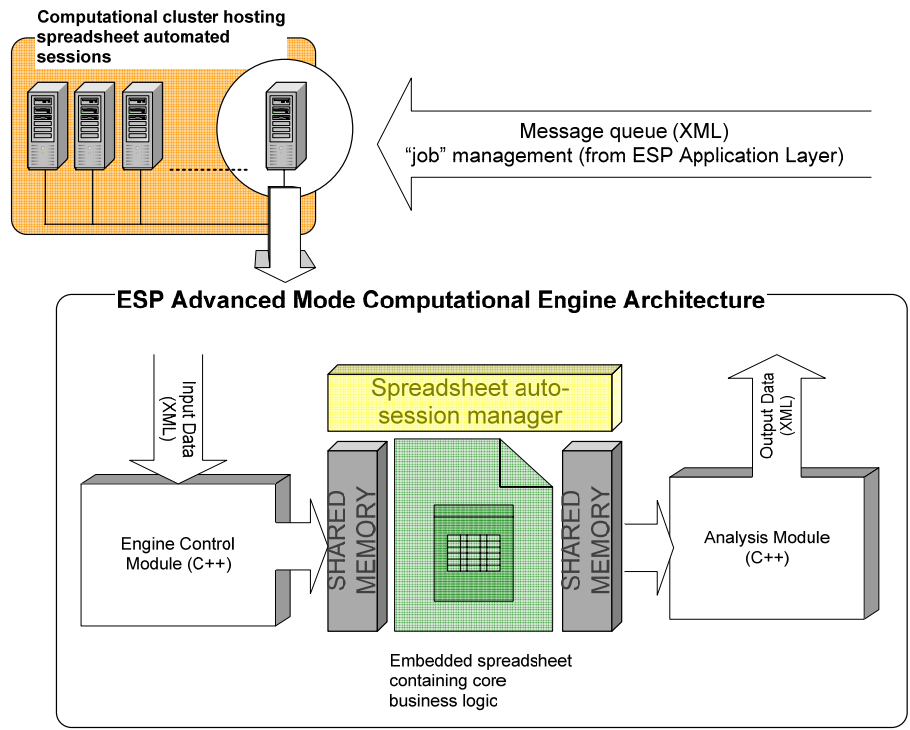

Figure 3. ESP Advanced Mode. The auto-spawned spreadsheet session is embedded within a (C++ coded) computational framework via shared-memory, with XML-based input/output data feeds to/from the Application Layer via queued-messaging.

Under this arrangement, various aspects of spreadsheet risk are eliminated. For example with specific regard to the sample application presented:

1. Complex, numerically-intensive Monte Carlo calculations are not subjected to the risk of being programmed into the spreadsheet by business analysts. Only the core deal logic is programmed in Excel by the business analyst.

2. Complex numerical method algorithms (e.g. matrix computations, central to many financial applications) are not subjected to the risk of being programmed in Excel (or VBA). Rather, best-of-breed compiled libraries (such as LAPACK for matrix computations) can be linked in via the ESP Advanced Mode interface.

3. In the specific case of Monte Carlo simulation, the use of pseudo-random numbers is central to the technique. In Excel, the programmer has no control over the "seed" of the pseudo-random number generator. Hence, it is impossible to replicate the outputs for a given set of inputs e.g. during the testing phase of the simulator. By contrast, under ESP, non-Excel-based pseudo-random number generators can be employed. These have the benefit of providing full control over the "seed", so that input/output replication can be realised for test cases. This eliminates the significant risk associated with not being able to properly test the simulator before deployment.

4. Since the ESP computational and data management framework is separate from the Excel application, proper runtime monitoring can be invoked. For example, if the spawned Excel session "hangs" (unfortunately, an all-to common occurrence, especially for large computations), the ESP framework can detect this, and, if necessary, shut-down the spreadsheet session, and inform the user that the computation has not proceeded successfully. This eliminates the operating risk associated with accepting the results from possibly incomplete spreadsheet computations.

5. Spreadsheet applications are notoriously slow at computation compared with compiled code. This is particularly evident when attempting to construct complex Monte Carlo simulations in Excel. By contrast, under the ESP computational framework, where the bulk of the numerically-intensive computations are performed in the compiled C++ code and only the deal logic remains in Excel, the speed of computation can be optimized. For example, on a Monte Carlo simulation of a large commercial real estate deal on high-end PC hardware, the ESP simulation framework is typically hundreds of times faster than the same simulation programmed in Excel, operating on the same Excel deal logic. By providing such performance advantages, the risks associated with the users' temptation to "run just a few Monte Carlo iterations" (i.e. to speed-up the Excel-based simulations) is mitigated. This can be significant, especially when calculating Probability-of-Default (PD) and Loss-Given-Default (LGD) for Basel II, where statistically-significant results can only be achieved with large numbers of Monte Carlo iterations (especially for low-risk deals). When running fewer iterations, the results can be wholly misleading.

6. With the separation between data and the spreadsheet, ESP enables any important parameters (e.g. number of Monte Carlo iterations, discussed in the previous item, macro-economic assumptions, centrally-set deal parameters such as "haircuts", etc) to be "locked down" such that the computations are performed consistently across the portfolio.

Although we have intentionally presented a complex application example (i.e. involving Monte Carlo iterations) to highlight the key benefits of the approach, it should be noted that *any* Excel spreadsheet can be configured for use under ESP Advanced mode. This is facilitated in a straightforward manner by the inclusion of an "add-in" (which contains the code hooks to the shared-memory interface) plus two specific worksheets, in an agreed format, which comprise the list of input and output variables, respectively.

## 4. CONCLUDING REMARKS

We have presented Risk Integrated's Enterprise Spreadsheet Platform (ESP), our approach to the near-elimination of spreadsheet risk in the enterprise computing environment. With a complex example application, Monte Carlo simulation, we have demonstrated how ESP provides a secure, robust, computational framework, yet with spreadsheets remaining at the core, thereby preserving the flexibility demanded by business analysts.

## 5. REFERENCES

SOX was introduced by the SEC to protect the public from accounting error and fraud in the wake of the Enron, WorldCom, and Arthur Andersen scandals. For one, it demands companies to archive for five years all business records including electronic records and emails. Additionally, each company's external auditors are required to audit and report on the internal control reports of management, submitting an annual report of the effectiveness of their internal accounting controls to the SEC. This affects public U.S. companies and non-U.S. companies with a U.S. presence.

Basel II defines operational risk as the risk of loss resulting from inadequate or failed internal processes, people or systems, or from external events. In the U.S., advanced measurement approaches are being approved by the government which factor in operational risk into the calculation of total capital requirements. The more effective a bank's internal operational risk management system is, the less money it needs to set aside in reserve. A bank needs to able to measure the risk and manage it with tools sensitive enough to analyze all internal and external data and perform scenario analysis. They also need to take into account any mitigating factors, such as insurance coverage. Central to Basel II compliance, and to a bank's ability to ultimately drive down its capital requirements, is the need to create a so-called "risk culture," one that recognizes there are many consequences for failing to handle information correctly.

The Banker, (2006), "Spreadsheet Risk", May 2006, pages 130-131.

Gartner, Inc., (2006), "Magic Quadrant for Business Intelligence Platforms, 1Q06."

A computational framework: Jafry, Sidoroff & Chi

Blank page